\documentclass[12pt,a4paper]{article}
\usepackage{graphicx}
\begin{document}
\begin{center}
{\large $R$ value measurements for $e^+e^-$
annihilation at 2.60, 3.07 and 3.65 GeV}\\
\end{center}

\noindent{\small{ M.~Ablikim$^{1}$,              J.~Z.~Bai$^{1}$,
Y.~Bai$^{1}$, Y.~Ban$^{11}$, X.~Cai$^{1}$, H.~F.~Chen$^{16}$,
H.~S.~Chen$^{1}$,              H.~X.~Chen$^{1}$, J.~C.~Chen$^{1}$,
Jin~Chen$^{1}$
% ...
{\center~~~~~~~~~~~~~~~~~~~~~~~~~~~~~~~~~~~~~~~~~~~~~~ (BES Collaboration)} \\
\vspace{0.2cm}
$^{1}$ Institute of High Energy Physics, Beijing 100049, People's Republic of China\\
% ...
$^{11}$ Peking University, Beijing 100871, People's Republic of China\\
% ...
$^{16}$ University of Science and Technology of China, Hefei 230026,
People's Republic of China\\
% ...
}

\date{\today}

\begin{abstract}
Using a data sample with a total integrated luminosity of 10.0
pb$^{-1}$ collected at center-of-mass energies of 2.6, 3.07 and 3.65
GeV with BESII, cross sections for $e^+e^-$ annihilation into
hadronic final states ($R$ values) are measured with statistical
errors that are smaller than $1\%$, and systematic errors that are
about $3.5\%$. The running strong interaction coupling constants
$\alpha_s^{(3)}(s)$ and $\alpha_s^{(5)}(M_Z^2)$ are determined from
the $R$ values.
\end{abstract}

\section{Introduction}

~~~~~The $R$ ratio is defined as the lowest level hadronic cross
section normalized by the theoretical $\mu^+\mu^-$ production cross
section in $e^+e^-$ annihilation
\begin{equation}\label{Rvalueq}
R=\frac{ \sigma_{had}^0(e^{+}e^{-}\to \gamma^{*}\to {\rm hadrons})}
{ \sigma_{\mu\mu}^0(e^{+}e^{-}\to \gamma^{*}\to \mu^{+}\mu^{-})},
\end{equation}
and it is an important input parameter for precision tests of the
Standard Model (SM). The errors on $R$ value measurements below 5
GeV have a strong influence on the uncertainties of the calculated
QED running electromagnetic coupling constant $\alpha(s)$, the muon
anomalous magnetic moment $(g-2)$ and global SM fits for the Higgs
mass \cite{smtest1,smtest2,smtest3}. In addition, precision
measurements of $R$ values between 2.0 and 3.7 GeV provide a test of
perturbative QCD and QCD sum rule calculations
\cite{pdg08,davier1,davier2}.

In 1998 and 1999, $R$ value measurements were made at 91 energy
points \cite{besr98,besr99} between 2 and 5 GeV by the BESII
\cite{BESII} experiment. The $R$ values were determined from the
expression
\begin{equation}
R_{exp}=\frac{ N^{obs}_{had} - N_{bg} } { \sigma^0_{\mu\mu} L
\epsilon_{trg} \epsilon_{had}^0 (1+\delta_{obs})}, \label{Rexp}
\end{equation}
where $N_{had}^{obs}$ is the number of observed hadronic events,
$N_{bg}$ is the number of QED background events surviving hadron
selection ($e^+e^-$, $\mu^+\mu^-$, $\tau^+\tau^-$, $\gamma\gamma$,
etc.), $L$ is the integrated luminosity, $\varepsilon_{trg}$ is the
trigger efficiency for hadronic events, $\varepsilon_{had}^0$ is the
hadronic efficiency without the simulation of initial state
radiation (ISR), and $(1+\delta_{obs})$ is the effective ISR
correction which includes the effect of radiated photons on the
hadronic acceptance. The average statistical errors of the $R$
values are $2 - 4\%$, and the systematical errors are $5 - 8\%$
depending on the energy point; for the latter, the errors associated
with the event selection, hadronic efficiency, and luminosity are
dominant.

In 2004, large-statistics data samples were accumulated at
center-of-mass energies of 2.60, 3.07 and 3.65 GeV; the total
integrated luminosity was 10.0 pb$^{-1}$. An additional
$65.2$ nb$^{-1}$ was accumulated at 2.2 GeV for the purpose of
tuning the parameters of the hadronic event generator. Improvements
in the event selection, tuning of generator parameters and
luminosity measurement have been made in order to decrease the
systematic errors. The previously used EGS-based (Electron Gamma
Shower) \cite{EGS} detector simulation (where hadronic interactions
were parametrized but not simulated) was replaced by a GEANT3-based
one \cite{GEANT3A,GEANT3B}, where both electromagnetic and hadronic
interactions are simulated. The consistency between data and Monte
Carlo (MC) has been validated using many high purity physics
channels \cite{SIMBES}. With these improvements, the errors on the
new measured $R$ values are reduced to about $3.5\%$.

In this letter, improvements compared to the previous measurements
are described, and the error analysis and results are reported.
Finally, the strong running coupling constant $\alpha_s^{(3)}(s)$
and $\alpha_s^{(5)}(M_Z^2)$ are determined from the $R$ values.

\section{Data analysis}

~~~~~The analysis used for this work is similar to that used in the
previous BESII $R$ measurements \cite{besr98,besr99}. Two large
sources of error in the measurement arise from the event selection
and the determination of the hadronic efficiency; these are strongly
correlated.

\subsection{Selection of hadronic events}

~~~~~In the BEPC energy region, data collected include processes
that originate from beam-beam collisions, $e^+e^-\to e^+e^-$,
$\mu^+\mu^-$, $\tau^+\tau^-$, $\gamma\gamma$, $e^+e^- X$ ($X$ means
any possible final state), and hadrons (including continuum and
resonant states), as well as beam associated backgrounds. The
observed final state charged particles are $e$, $\mu$, $\pi$, $K$,
and $p$. To test the hadron selection criteria, described below,
different types of backgrounds are selected using specialized
criteria, and most of them are rejected with good efficiency
\cite{besr98,besr99,evtselimprv}.

The candidate hadronic events are classified by their number of
charged tracks. The selection of hadronic events is done in two
successive steps: one at the track level and the other at the event
level.

(I) Track level: only charged tracks that are well fitted to a helix
are considered; the point of closest approach of the track,
signified by $(V_x, V_y, V_z)$, must be within $2$ cm of the
beam-line in the $x$-$y$ plane with no restriction on $V_z$; the
angle between the track and the $z$ axis is limited by the coverage
of the main drift chamber (MDC) to be within $|\cos\theta|<0.84$; a
momentum cut $p<E_b(1+0.1\sqrt{1+E_b^2})$ removes tracks with
unphysically high momentum ($E_b$ is the beam energy in units of
GeV); a time of flight (TOF) requirement $t_{TOF}\leq t_p+2$ ns is
applied to reject events with unphysical times ($t_p$ is the
expected value for a proton with momentum $p$); the energy deposited
in the barrel shower counter (BSC), $E_{BSC}$, must be less than the
minimum of 1 GeV and $0.6 E_B$; and the number of hit layers in the
muon counter must be smaller than 3.

A neutral cluster in the BSC is considered to be a photon candidate
if the angle between the nearest charged track and the cluster is
greater than $25^\circ$; the difference between the angle of the
cluster development direction and the photon emission direction in
the BSC is less than $30^\circ$; and the number of hit layers in the
BSC is larger than two.

(II) Event level: the requirement that the total deposited energy,
$E_{BSC}^{sum}$, is greater than the maximum of 0.5 GeV and
$0.28E_b$
eliminates most of the beam-associated backgrounds.  The selected
tracks must not all point into the forward ($\cos \theta > 0$) or
the backward ($\cos \theta < 0$) hemisphere.  For multi-track events
($n_{ch}\geq3$), no further requirement is used. For two-track
events, the two charged tracks cannot be back-to-back, so the angle
between them is required to be less than $165^\circ$, and there must
be at least two isolated photons with energy $E_\gamma>0.1$ GeV that
are well separated from charged tracks, {\it i.e.} the distance
between the neutral and charged tracks at the first layer of the BSC
must be larger than 34 cm in the $x$-$y$ plane, or larger than 60 cm
in the $z$ direction.

The above described selection criteria for hadronic events with
$n_{ch}\geq2$ are almost the same as used in the previous
measurements \cite{besr98,besr99}. In the BEPC energy region, the
number of events with one observed/reconstructed charged track in
BESII accounts for about $8$-$13\%$ of all hadronic events. The
omission of one-track (and zero-track) hadronic events introduces
some uncertainty in the tuning the parameters of the hadronic event
generator parameters; this in turn induces a sizable systematic
error in the hadronic efficiency. However, for single-track events,
contamination from beam-associated backgrounds is significant.
Therefore, a more restrictive hadronic event selection is applied
\cite{evtselimprv}: the event must have one charged track that is
well fitted to a helix (the event can have additional any number of
charged tracks with poor helix fits). If this charged track is
identified as $e^\pm$, the event is rejected. If its momentum is in
the range $p>1$ GeV and the number of hit layers in the muon counter
is larger than 1, the event is also rejected. For each single-track
event, the number of photons with energy $E_\gamma>0.1$ GeV should
be two or more. To further suppress background from fake photons,
only events with one well fitted charged track and at least one
reconstructed $\pi^0$ are considered as single-track hadronic
events. A 1-C fit
is applied under the $\pi^0\to\gamma\gamma$ hypothesis. For
candidates with more photons ($n_\gamma\geq3$), the $\gamma\gamma$
pair combination with the smallest $\chi^2$ is chosen.  The $\chi^2$
probability for the 1-C fit is required to be larger than $1\%$.
Figure \ref{datalundarea}(h) shows the invariant $\gamma\gamma$ mass
distribution for one-track events satisfying requirements for data
and MC (the MC is normalized to the integrated luminosity of the
data). The good agreement indicates that the rate of one-track
events in data and MC are commensurate, and the residual
beam-associated background is small.

For selected events, the weighted average vertex position,
$\overline{V}_z$, is determined using
\begin{equation}
\overline{V}_z=\frac{\sum_{i=1}^{n}
V_z(i)/\tilde{\chi}_i^2}{\sum_{i=1}^{n} 1/\tilde{\chi}_i^2},
\end{equation}
where $V_z(i)$ is the $z$ coordinate of the $i$th selected track,
$\chi_i^2$ is the track fitting chi-square, $\tilde{\chi}_i^2
=\chi_i^2/n_{D.O.F.}$, $n_{D.O.F.}$ is the number of degrees of
freedom, and $n$ is the number of tracks in the event.  Figure
\ref{vzdis} shows the $\overline{V}_z$ distributions for candidate
hadronic events (including the residual beam-associated and QED
backgrounds) at 2.6, 3.07 and 3.65 GeV.  Signal events produced by
$e^+e^-$ collisions originate near the collision point (in the
neighborhood of $z=0$), and the non-beam-beam backgrounds, such as
those from beam-gas and beam-wall scattering, are distributed all
along the beam direction (In order to show the shape of the
$\overline{V}_z$ distribution of residual beam-associated
backgrounds clearly, a logarithmic vertical scale is used). The
number of observed hadronic events $N_{had}^{obs}$ is determined by
fitting the $\overline{V}_z$ distribution with a Gaussian to
describe the hadronic events and an $m$-th degree polynomial for the
residual beam associated backgrounds.


The numbers of residual QED background events,  $N_{bg}$ in
Eq.(\ref{Rexp}), are determined from MC simulations using QED event
generators with an accuracy of $1\%$ \cite{qedgenerator}, where
\begin{equation}
N_{bg}=L[\epsilon_{ee}\sigma_{ee}+\epsilon_{\mu\mu}\sigma_{\mu\mu}
+\epsilon_{\tau\tau}\sigma_{\tau\tau}+\epsilon_{\gamma\gamma}\sigma_{\gamma\gamma}].
\end{equation}
Here $\sigma_{ee}$ is the production cross section for Bhabha events
given by the corresponding generator, $\epsilon_{ee}$ is the
efficiency for Bhabha events that pass the hadronic event selection
criteria, and other symbols have corresponding meanings. The values
of $\epsilon_{ee}$ and $\epsilon_{\mu\mu}$ are about
$5\times10^{-4}$, and $\epsilon_{\tau\tau}$ is $36.45\%$ at 3.65
GeV. The errors on $N_{bg}$ are given in Table \ref{Rerror}.  The
amount of background from $e^+e^-\to e^+e^-X$ that survives hadron
selection is much smaller than $1\%$ of $N_{bg}$ and is neglected.

\subsection{Tuning the LUARLW parameters}

~~~~~The hadronic efficiency is determined using the LUARLW hadronic
event generator \cite{luarlwhu}. The physical basis of LUARLW is the
Lund area law \cite{luarlwbohu}. The production of hadrons is
described as the fragmentation of a semi-classical relativistic
string \cite{bobook}, and the quark components of the string and
decays of unstable particles are handled by subroutines in JESTSET
\cite{jetset,jetset1,jetset2}.

Both generators LUARLW and JETSET have some phenomenological
parameters that have to be determined from data. The important
parameters in JETSET are $PARJ(1-3)$, which are responsible for the
suppression of diquark-diantiquark pair production, quark pair
production, and the extra suppression of strange diquark production
respectively, and $PARJ(11-17)$, which control the relative
probabilities of different spin mesons. The values of these
parameters have been tuned with LEP data, as well as with
information from lower energy $e^+e^-$ colliders \cite{pr29197}. The
generator LUARLW has also been tuned with BES data taken at 2.2,
2.6, 3.07 and 3.65 GeV. The tuned parameters for the Lund area law
are mainly the dynamical parameter $b$ and those related to the
initial multiplicity distributions (including neutral clusters and
charged tracks) from the string decay \cite{luarlwhu}.

The basic method is to find a set of parameters that make various
distributions (especially those related to the hadronic selection)
simulated by MC agree well with experimental data at all of the
measured energy points. The distributions used for the data$-$MC
comparison are: the multiplicities of charged tracks and neutral
clusters, the $V_{x-y}$ and $V_z$ coordinates of charged tracks, the
charged track momentum, the polar-angle $\theta$ between tracks and
the beam direction, the deposited energy in the BSC, the time of
flight, and fractions of $\pi^\pm$, $K^\pm$, and some other
short-lived particles ($\pi^0$, $K_S$, $\phi$, $\Lambda$), etc..
With these distributions, the systematic errors corresponding to
each criteria used in the hadronic event selection can be
determined. Figure \ref{datalundarea} shows, for example, some
comparisons between data and the LUARLW MC at 3.65 GeV, where
reasonable agreement is evident. More distributions at other energy
points are also compared, which can be found in Refs.
\cite{evtselimprv,beamgassep}.

The parameters for detector simulation are studied using data. The
constant files for dead and hot channels, and the detailed detector
responses are inputs for detector simulations.


\subsection{Trigger, luminosity and ISR}

~~~~~The trigger conditions are almost the same as those used for
the $R$ measurements reported in Refs.~\cite{besr98,besr99}, and the
details about the values and errors of the trigger efficiency are
described in Refs.~\cite{trigger}. Since one-track hadronic events
are also included in this measurement, the TOF back-to-back hit
trigger requirement is not used, thereby making the trigger
conditions somewhat loser than before. The trigger table used in
data taking is given in Ref. \cite{evtselimprv}. The trigger
efficiencies $\epsilon_{trg}$ for hadronic events is determined as
$99.8\%$, and its associated errors is conservatively estimated to
be $0.5\%$.

The integrated luminosity $L$ is measured with wide-angle
($|\cos\theta|<0.6$) Bhabha events. The measurement method is
similar to that described in Refs.\cite{besr98,besr99}.
The Bhabha events are selected using only BSC information, and the
simulation of the BSC is significantly improved by the package
(SIMBES) described in \cite{SIMBES}, which provides better
consistency between MC and data. In addition, the BSC selection
efficiencies, determined by simulation, are corrected using
correction factors determined with another Bhabha sample, that is
selected using only MDC information. The efficiency correction
factors range from $0.994$ to $1.026$ with an uncertainty of about
1.4\% for the different energy points. In addition, the contribution
from $e^+e^-\to\gamma\gamma$ is subtracted explicitly. As a result,
the precision of the luminosity determined is significantly
improved, and the systematic uncertainties are about 2\%. For
example, at 2.6 GeV the systematic error is 1.9\%, of which the
trigger efficiency contributes 0.5\%, the MC generator 0.5\%, event
selection 1.2\%, and the Bhabha correction 1.3\%.

An $\mathcal{O}(\alpha^3)$ Feynman-diagram-based calculation for the
initial state radiative (ISR) correction is used in both the
calculation of the ISR factor $(1+\delta_{obs})$ and the simulation
of radiative events by LUARLW. A detailed description of the ISR
treatment can be found in Refs.
\cite{cb4160,cb5160,isrhuhm,bonneau,berends}. In the ISR simulations
and calculations, the contributions from both continuum and
resonances are considered. The inclusive hadronic cross section
below 5 GeV uses the experimental value, and above 5 GeV uses the
theoretical value predicted by QCD \cite{pdg06}. The quantities
related to the narrow $J/\psi$ and $\psi'$ are treated analytically.

Figure \ref{radeff} shows the detection efficiencies for hadronic
events simulated with LUARLW for the cases where an initial-state
$e^+$ or $e^-$ radiates a photon with energy fraction $k\equiv
E_\gamma/E_{b}$ and those events with $n_{ch}\geq1$ are selected.
Curves fitted to $\epsilon(k)$ are used in the calculation of
$(1+\delta_{obs})$ at different energies. The values of
$(1+\delta_{obs})$ and their errors are listed in Tables
\ref{Rvalue} and \ref{Rerror}, respectively.

\section{Error analysis}

~~~~~The Feynman-diagram-based ISR simulated angle and momentum
distributions for the radiated photon are built into LUARLW, which
allows the hadronic efficiency $\bar{\epsilon}_{had}$ including
radiative effects to be obtained, averaging over the ISR spectrum.
The number of hadronic events $N_{had}^{obs}$, the hadronic
efficiency $\bar{\epsilon}_{had}$, and their errors are correlated.
The equivalent number of hadronic events, which corresponds to the
number of hadronic events produced at the collision point, is
defined as
\begin{equation}
N_{had}=\frac{N_{had}^{obs}}{\bar\epsilon_{had}}\label{nhadgen}.
\end{equation}
The combined systematic error associated with the event selection
and hadronic efficiency is denoted as $\Delta N_{had}$. This error
is caused by discrepancies between data and MC samples for the
hadronic selection criteria discussed in Section 2.1. In addition,
an uncertainty associated with the parameters of the MC
hadronization model is estimated to be about $1\%$ by comparing
different sets of tuned parameters, and is included in the error of
the hadronic event efficiency.

The error on $N_{had}^{obs}$ due to the choice of the degree of the
polynomial used in the fitting is less than $0.7\%$.
The fit errors for $N_{had}^{obs}$, which are calculated from the
uncertainties in the fitted parameters of the Gaussian signal peaks,
are $1.34\%$ at 2.6 GeV, $1.11\%$ at 3.07 GeV, and $0.73\%$ at 3.65
GeV. The total $\Delta N_{had}$ is the quadratic sum of all
fractional errors.

The uncertainty in the effective ISR factor $(1+\delta_{obs})$ due
to errors of the hadronic cross sections at the different effective
energies for radiative events is considered (the errors on the
hadronic cross section given in the PDG06 tables \cite{pdg06} are
used); these decrease with increasing energy from $0.9\%$ to
$0.1\%$. For comparison, another approach based on structure
functions \cite{kureav} is also used at all energy points. The
differences in the values of $(1+\delta)$ for the two schemes are
smaller than $1.1\%$.

A conclusion of the KLN theorem is that the radiative corrections
due to final state radiation (FSR) are negligible for a measurement
of the inclusive hadronic cross section that sums over all hadronic
final states \cite{tsai}. At the present level of precision, the FSR
correction factor in Eq.(\ref{Rexp}) can be neglected. However, the
absence of final state radiation in the event generator introduces
some error into the determination of the hadronic event detection
efficiency. The masses of the produced hadrons in the final states
are much greater than that of the initial radiative $e^\pm$. As a
result, the effect of FSR is much weaker than initial
bremsstrahlung. Its influence is estimated to be $0.5\%$ and is
included in the error.

The 0-track hadronic events are not selected in this analysis, and
the influence of 0-track events on the parameter tuning of LUARLW is
not considered. This introduces some error into the hadronic event
efficiency determination. Events with no charged tracks cannot be
well separated from background. The fraction of 0-track events is
estimated from the MC to be $3.4\%$ at 2.6 GeV, $2.9\%$ at 3.07 GeV,
and $2.4\%$ at 3.65 GeV. If the difference for 0-track events
between MC and data is conservatively assumed to be $20\%$, the
errors for the lost/unobserved 0-track events are $0.7\%$, $0.6\%$
and $0.5\%$, respectively. The error related to 0-track events is
included into the error of $N_{had}$ defined in Eq.~(\ref{nhadgen}).

In this analysis, hadronic events are classified according to their
number of charged tracks. Therefore, errors in the tracking
efficiency $\sigma_{trk}$, the differences in the track
reconstruction efficiency between data and MC, introduce some error
into the classification and counting of the number of events. For an
event with $n_{ch}$ charged tracks, the probability that $n_{er}$ of
$n_{ch}$ tracks are wrongly constructed roughly obeys a binomial
distribution $B(n_{er};n_{ch},\sigma_{trk})$, where the parameter
$\sigma_{trk}\sim2\%$ is the tracking efficiency error at BESII. The
$R$ value measurement is, in fact, a counting of the number of
hadronic events, so only those cases where all $n_{ch}$ tracks in an
event are incorrectly reconstructed ($n_{er}=n_{ch}$) will introduce
an error into $\Delta N_{had}$. Considering the distribution of
charged multiplicity $P(n_{ch})$ for the inclusive hadronic sample
(such as shown in Figure \ref{datalundarea}(a)), the effective error
of tracking efficiency is
\begin{equation}\label{errortrk}
\Delta\epsilon_{trk}
=\sum_{n_{ch}\geq1}P(n_{ch})B(n_{er}=n_{ch};n_{ch},\sigma_{trk}).
\end{equation}
The estimated values of $\Delta\epsilon_{trk}$ are listed in Table
\ref{Rerror}. Since the fraction of single-track events decreases
with increasing center-of-mass energy, the error
$\Delta\epsilon_{trk}$ also decreases with energy.

The errors on the $R$ values are determined with Eq.~(\ref{Rexp})
(including all errors analyzed in last section and and estimated by
Eq.~(\ref{errortrk})). The final $R$ values are $2.18\pm0.02\pm0.08$
at 2.6 GeV, $2.13\pm0.02\pm0.07$ at 3.07 GeV, and
$2.14\pm0.01\pm0.07$ at 3.65 GeV, which are summarized in Table
\ref{Rvalue}.

As a cross check, the $R$ values are determined using the relation
\begin{equation}
R_{exp}=\frac{ N^{obs}_{had} - N_{bg} } { \sigma^0_{\mu\mu} L
\epsilon_{trg} \bar{\epsilon}_{had}(1+\delta)}, \label{Rexpt}
\end{equation}
where $\bar{\epsilon}_{had}$ is the hadronic efficiency averaged
over all of the ISR spectrum, and $(1+\delta)$ is the corresponding
theoretical ISR factor. The ISR scheme used to simulate
$\bar{\epsilon}_{had}$, $\epsilon^0_{had}$ and to calculate
$(1+\delta)$ and $(1+\delta_{obs})$ are the same in order to keep
the consistency between theoretical calculation and simulation. The
$R$ values determined with Eq.~(\ref{Rexpt}) are
$2.17\pm0.01\pm0.07$ at 2.6 GeV, $2.13\pm0.01\pm0.07$ at 3.07 GeV,
and $2.16\pm0.01\pm0.08$ at 3.65 GeV.  The mean $R$ values obtained
using Eqs.~(\ref{Rexp}) and (\ref{Rexpt}) are consistent to within
$1\%$.

A cross check is also made by selecting hadronic events with
$n_{ch}\geq2$ as was done in Refs.~\cite{besr98,besr99}. In this
case, the $R$ values at the three energy points are
$2.20\pm0.02\pm0.08$, $2.13\pm0.02\pm0.07$, and
$2.15\pm0.01\pm0.08$, respectively. The differences in the mean $R$
values determined by selecting hadronic events with $n_{ch}\geq1$
and $n_{ch}\geq2$ are consistent within $1\%$.

\section{Results and discussion}

~~~~~Tables \ref{Rvalue} and \ref{Rerror} list the quantities used
in the determination of $R$ using Eq.~(\ref{Rexp}) and the
contributions to the total error. The results are shown in
Fig.~\ref{besr04}, together with previous measurements. The errors
on the $R$ values reported here are about $3.5\%$. The $R$ values
are consistent within errors with the prediction of perturbative QCD
\cite{pdg08}.

Compared with our previous results \cite{besr98,besr99}, the
measurement precision has been improved due to three main
refinements to the analysis: (1) the simulation of BES including
both of the hadronic and electromagnetic interactions with a GEANT3
based package SIMBES that has a more detailed geometrical
description and matter definition for the sub-detectors; (2) large
data samples are taken at each energy point, with statistical errors
smaller than $1\%$; (3) the selected hadronic event sample is
expanded to include one-track events, which supplies more
information to the tuning of LUARLW, and results in the improved
values of parameters and hadronic efficiency.

In another BESII work, parts of the data sample taken at 3.65 GeV
with a luminosity of 5.536 pb$^{-1}$ and at 3.665 GeV with a
luminosity of 998.2 nb$^{-1}$ are used, the hadronic events with
more than 2-tracks ($n_{ch}\geq3$) are selected, and the averaged
$R$ value is $R=2.218\pm0.019\pm0.089$ which has an error of $4.1\%$
\cite{plb6412006145}.

Based on the $R$ values in this work and the perturbative QCD
expansion that computes $R_{QCD}(\alpha_s)$ to
$\mathcal{O}(\alpha_s^3)$~\cite{rqcdalphas}, the strong running
coupling constant $\alpha_s^{(3)}(s)$ can be determined at each
energy point \cite{kuhn,ralphas}. The obtained $\alpha_s^{(3)}(s)$
values are evolved to 5 GeV, and the weighted average of the
measurements $\bar\alpha_s^{(4)}$(25 GeV$^2$) is listed in Table
\ref{alphas}. When evaluated at the $M_Z$ scale, the resulting value
is $\alpha_{s}^{(5)}(M_Z^2) = 0.117^{+0.012}_{-0.017}$, which agrees
with the world average value within the quoted errors \cite{pdg08}.

\begin{table}[hbtp]
\begin{center}
\caption{Items used in the determination of $R$ at each energy
point.}\label{Rvalue}
\begin{tabular}{ccrrcccccc} \hline
$E_{cm}$(GeV) & $L$ ($pb^{-1}$) & ~$N^{obs}_{had}$~  & ~$N_{bg}$ &
~$\epsilon_{had}^0$ (\%)~ & ($1+\delta_{obs}$) & ~~$R$~~&
~~$\sigma_{sta}$ & $\sigma_{sys}$\\\hline
2.60 & 1.222 & 24026 & 193 & 63.81 & 1.08 & 2.18 & 0.02 & 0.08\\
3.07 & 2.291 & 33933 & 208 & 67.63 & 1.11 & 2.13 & 0.02 & 0.07\\
3.65 & 6.485 & 83767 & 4937& 71.83 & 1.21 & 2.14 & 0.01 & 0.07\\
 \hline
\end{tabular}
\end{center}
\end{table}

\begin{table}[hbtp]
\begin{center}
\caption{Summary of the systematic errors (\%).}\label{Rerror}
\begin{tabular}{ccccccccc} \hline
$E_{cm}$(GeV) & $L$ & $N_{had}$  & $N_{bg}$ & $\Delta\epsilon_{trk}$
& $\epsilon_{trg}$ & ($1+\delta_{obs}$) & Total\\\hline
2.60 & 2.00 & 2.79 & 0.05 & 0.32 & 0.50 & 1.18  & 3.68 &  \\
3.07 & 1.96 & 2.53 & 0.05 & 0.29 & 0.50 & 1.15  & 3.45 &  \\
3.65 & 1.38 & 2.74 & 0.35 & 0.26 & 0.50 & 1.10  & 3.33 & \\
 \hline
\end{tabular}
\end{center}
\end{table}

\vspace{-5mm}

\begin{table}[h]
\begin{center}
\caption{$\alpha_{s}(s)$ determined from $R$ values at 2.600, 3.070,
and 3.650 GeV, and evolved to 5 GeV. The first and second errors are
statistical and systematic, respectively. Shown in the last two
columns are the weighted averages of the three measurements at 5 GeV
and $M_Z$.}\vspace{2mm}
\begin{tabular}{ccccc}\hline
$\sqrt{s}$(GeV)& $\alpha_{s}^{(3)}(s)$  & $\alpha_{s}^{(4)}(25{\rm
GeV}^2)$~&$\bar\alpha_{s}^{(4)}(25{\rm GeV}^2)$ &$\alpha_{s}^{(5)}(M_z^2)$\\
\hline 2.60
&$0.266^{+0.030+0.125}_{-0.030-0.116}$&$0.212^{+0.018+0.068}_{-0.019-0.086}$ \\
3.07
&$0.192^{+0.029+0.103}_{-0.029-0.101}$&$0.169^{+0.022+0.074}_{-0.023-0.086}$ & $0.209^{+0.044}_{-0.050}$ &$0.117^{+0.012}_{-0.017}$\\
3.65
&$0.207^{+0.015+0.104}_{-0.015-0.104}$&$0.189^{+0.012+0.082}_{-0.013-0.091}$\\
\hline
\end{tabular}
\label{alphas}
\end{center}
\end{table}

The BES collaboration thanks the staff of BEPC and computing center
for their hard efforts. BES collaboration also thanks B. Andersson
for helping in the development of generator LUARLW during 1998-1999.
This work is supported in part by the National Natural Science
Foundation of China under contracts Nos. 19991480, 19805009,
19825116, 10491300, 10225524, 10225525, 10425523, 10625524,
10521003, the Chinese Academy of Sciences under contract No. KJ
95T-03, the 100 Talents Program of CAS under Contract Nos. U-11,
U-24, U-25, and the Knowledge Innovation Project of CAS under
Contract Nos. U-602, U-34 (IHEP), the National Natural Science
Foundation of China under Contract No. 10225522 (Tsinghua
University), and the Department of Energy under Contract No.
DE-FG02-04ER41291 (U. Hawaii).

\vspace{+10mm}

\begin{figure}[htbp]
\begin{center}
\includegraphics[width=5cm,height=5cm]{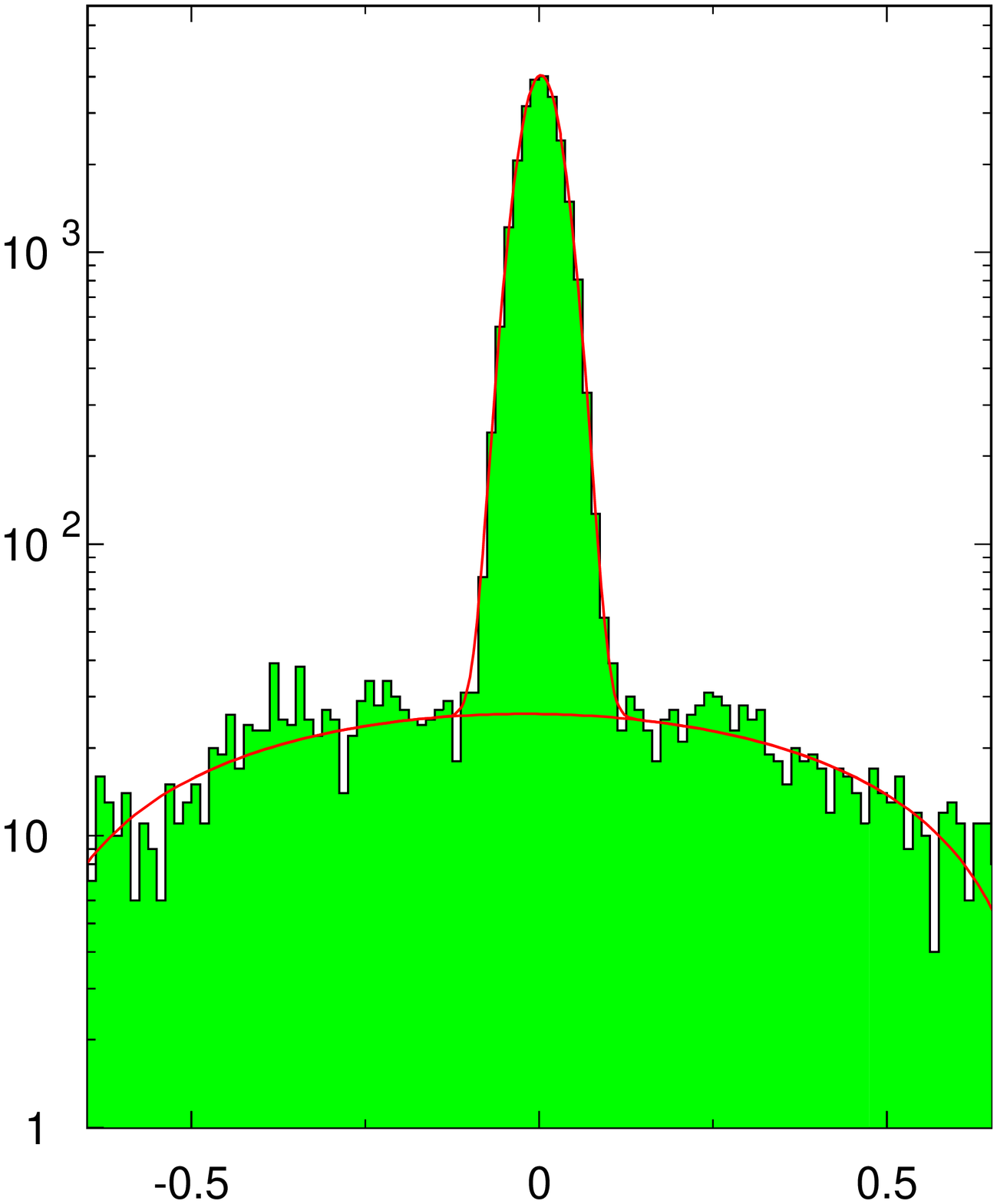}
\includegraphics[width=5cm,height=5cm]{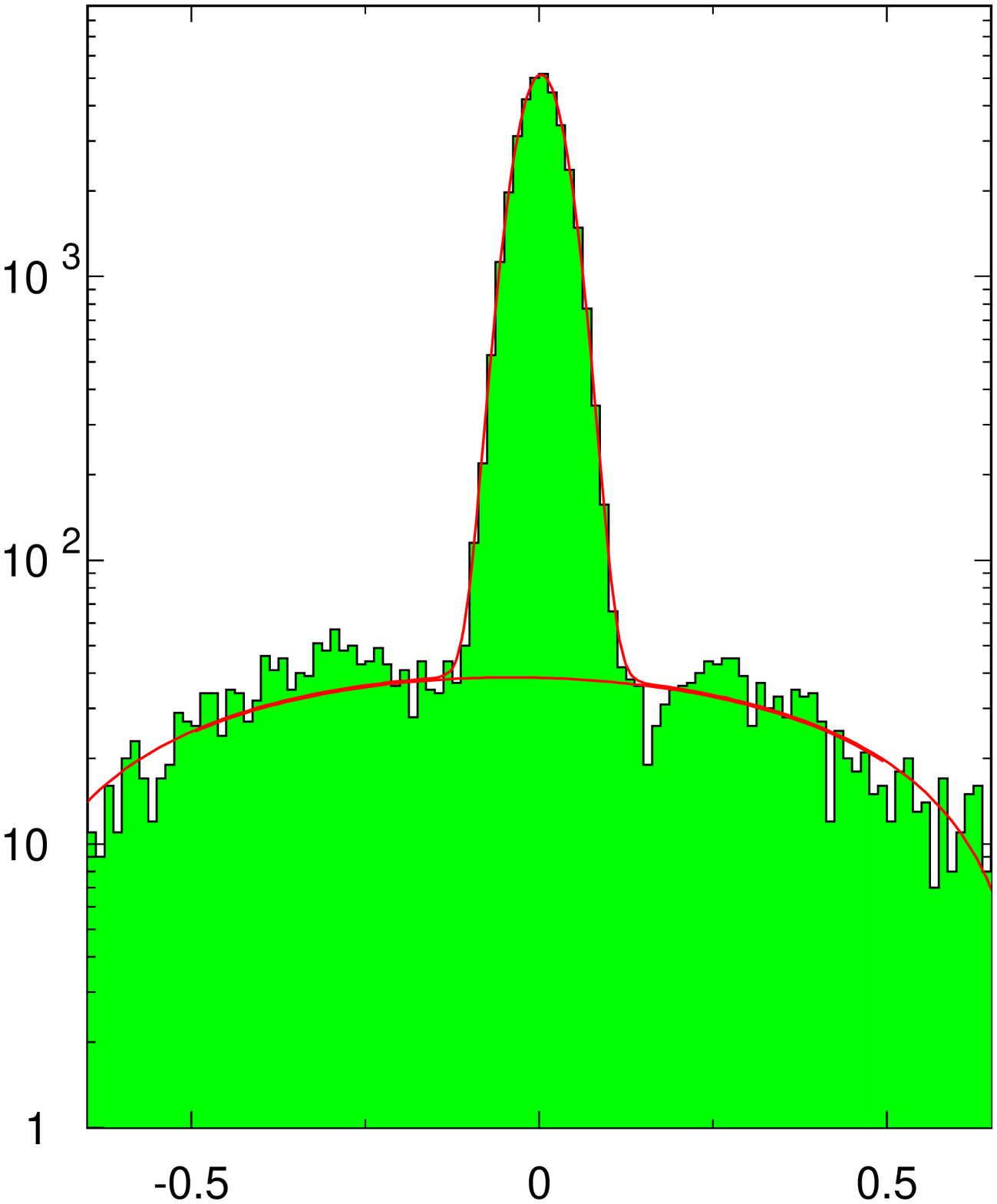}
\includegraphics[width=5cm,height=5cm]{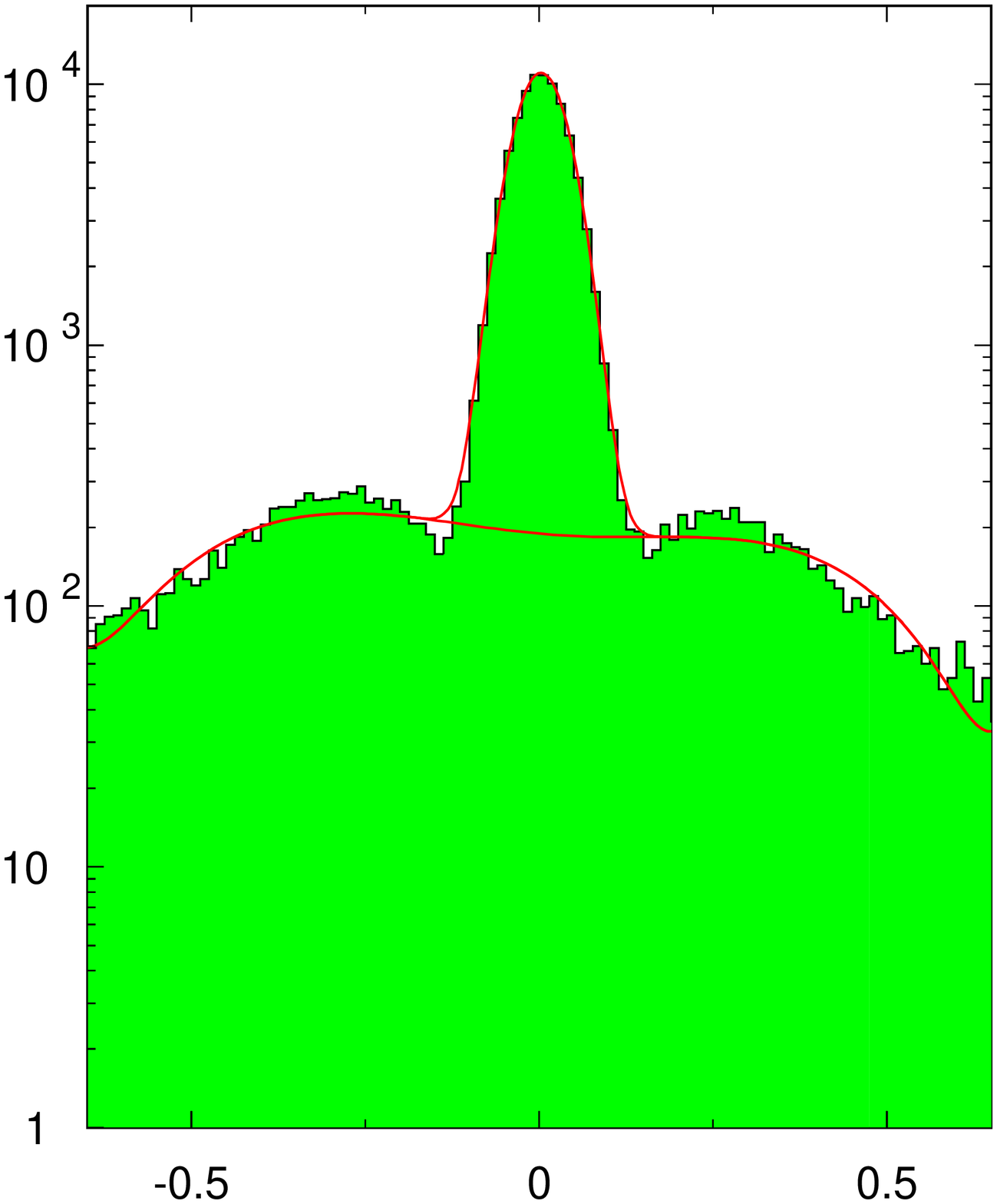}
\put(-345,125){2.60GeV}\put(-195,125){3.07GeV}\put(-50,125){3.65GeV}\caption{
Distributions of $\overline{V}_z$ for candidate hadronic events. The
distributions are fitted by a Gaussian to represent the signal and a
polynomial to describe the residual beam-associated
backgrounds.}\label{vzdis}
\end{center}
\end{figure}

\begin{figure}[htpb]
\includegraphics[width=3.80cm,height=3.85cm]{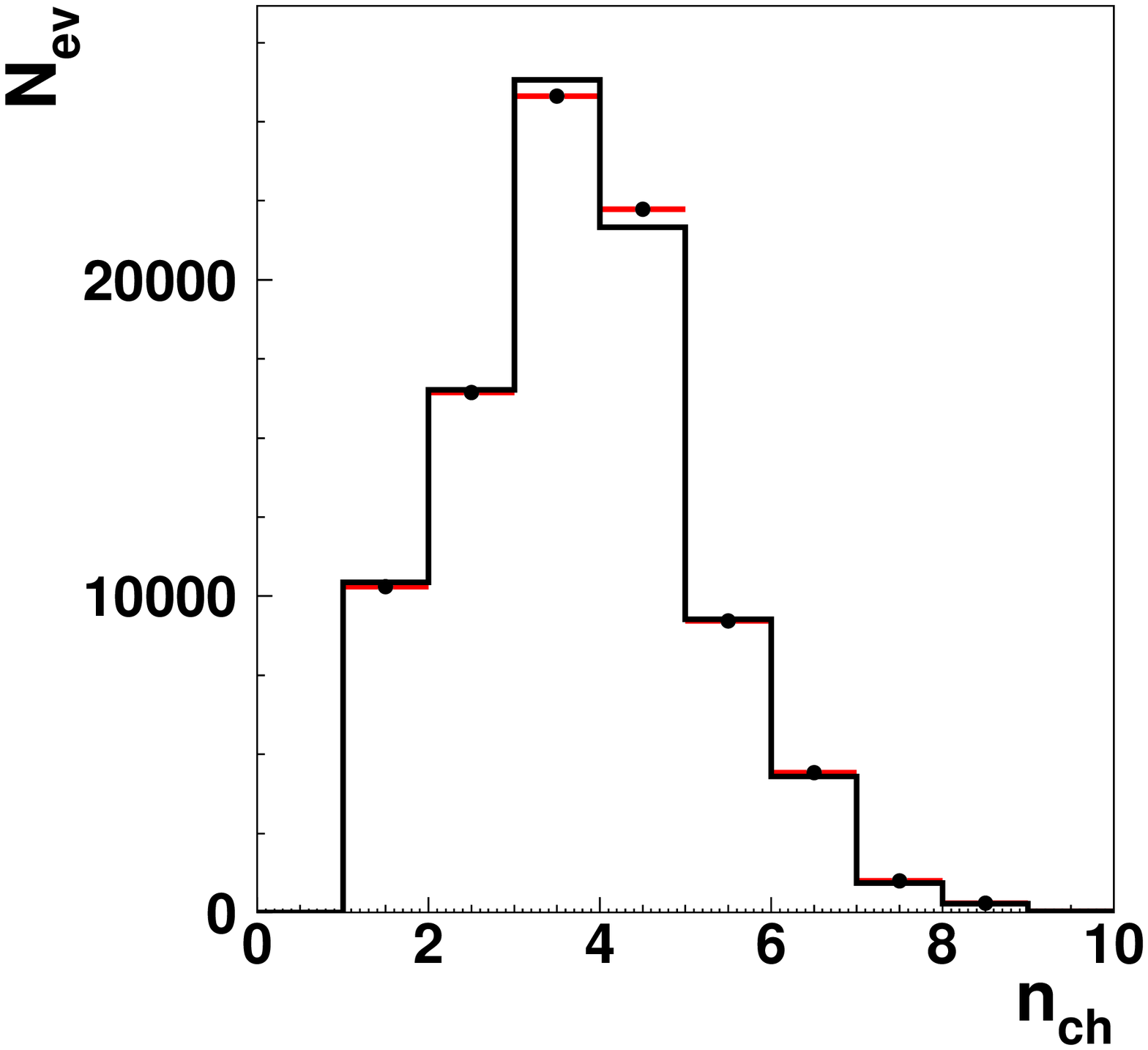}
\put(-30,95){(a)}\put(75,95){(b)}\put(195,95){(c)}\put(290,95){(d)}
\includegraphics[width=3.75cm,height=3.85cm]{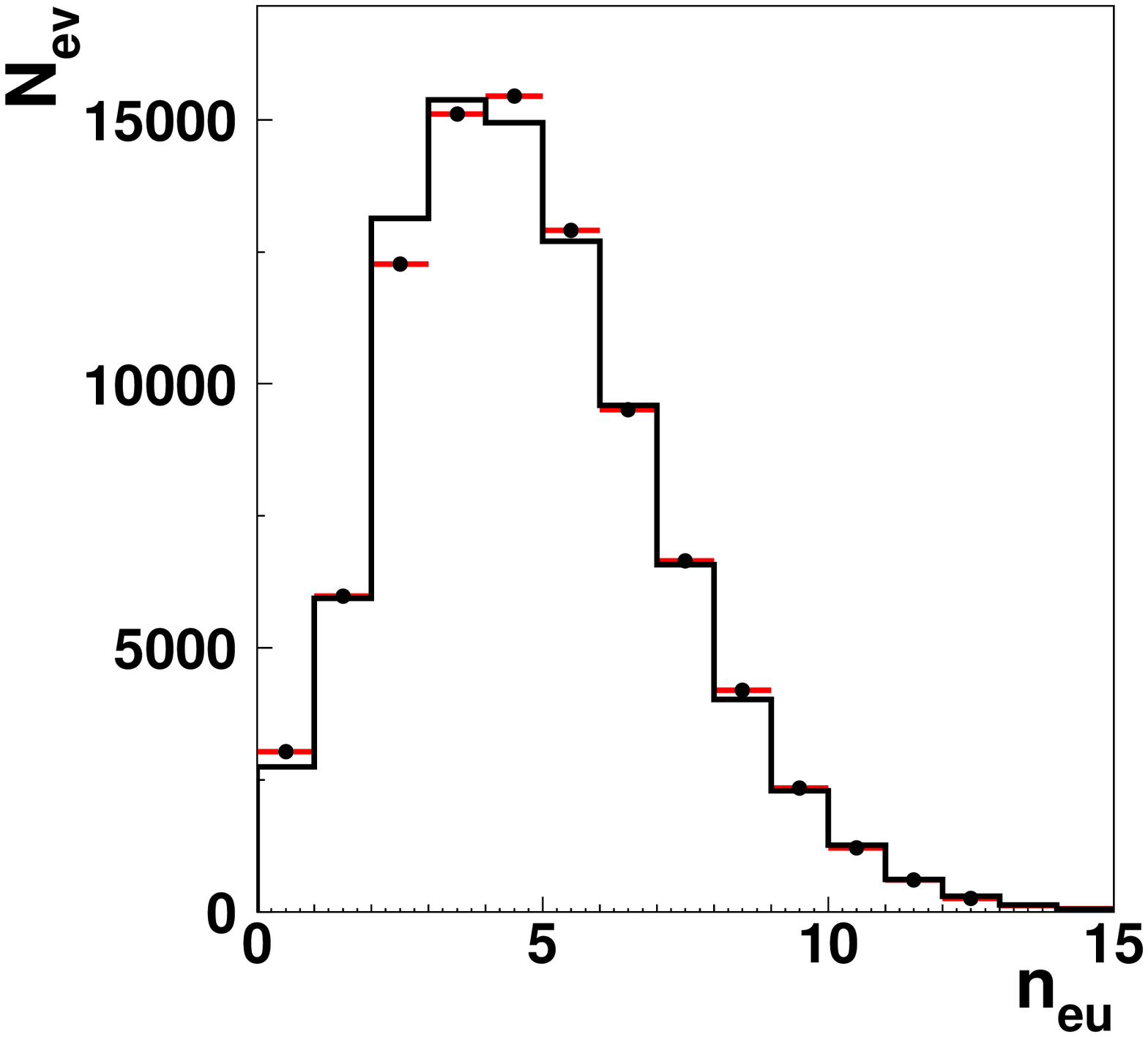}
\includegraphics[width=3.75cm,height=3.75cm]{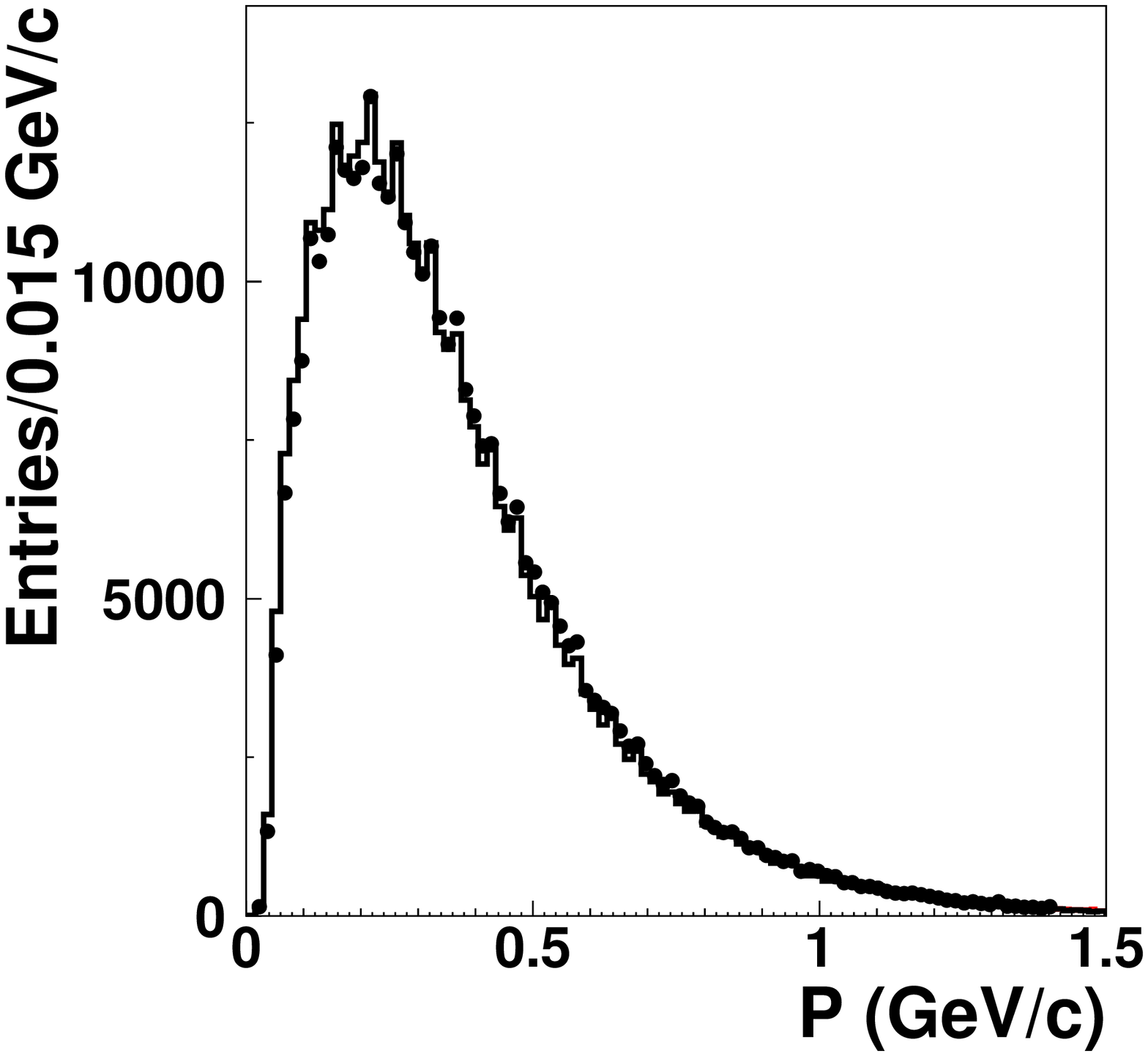}
\includegraphics[width=3.70cm,height=3.70cm]{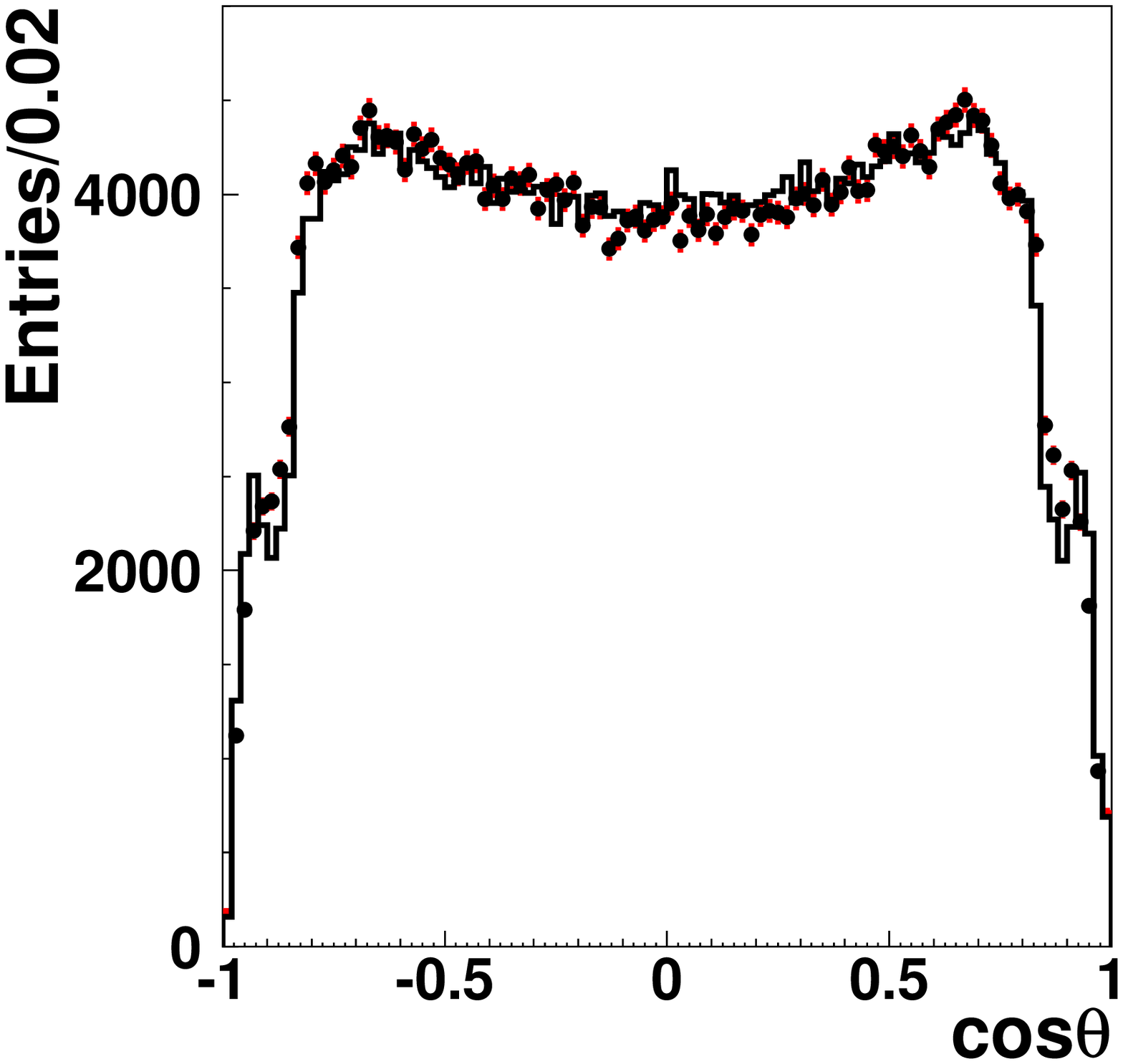}
\includegraphics[width=3.85cm,height=3.75cm]{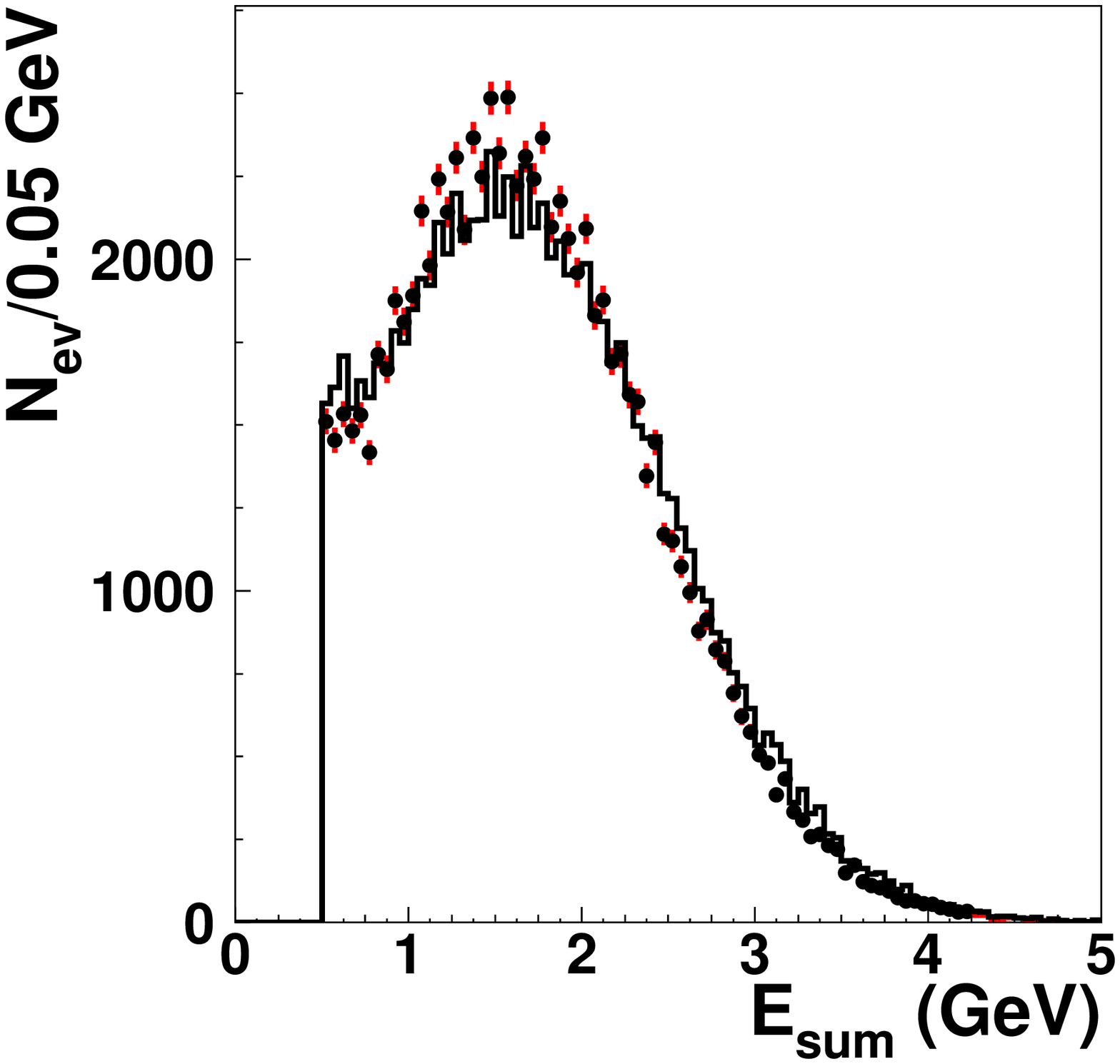}
\includegraphics[width=3.85cm,height=3.75cm]{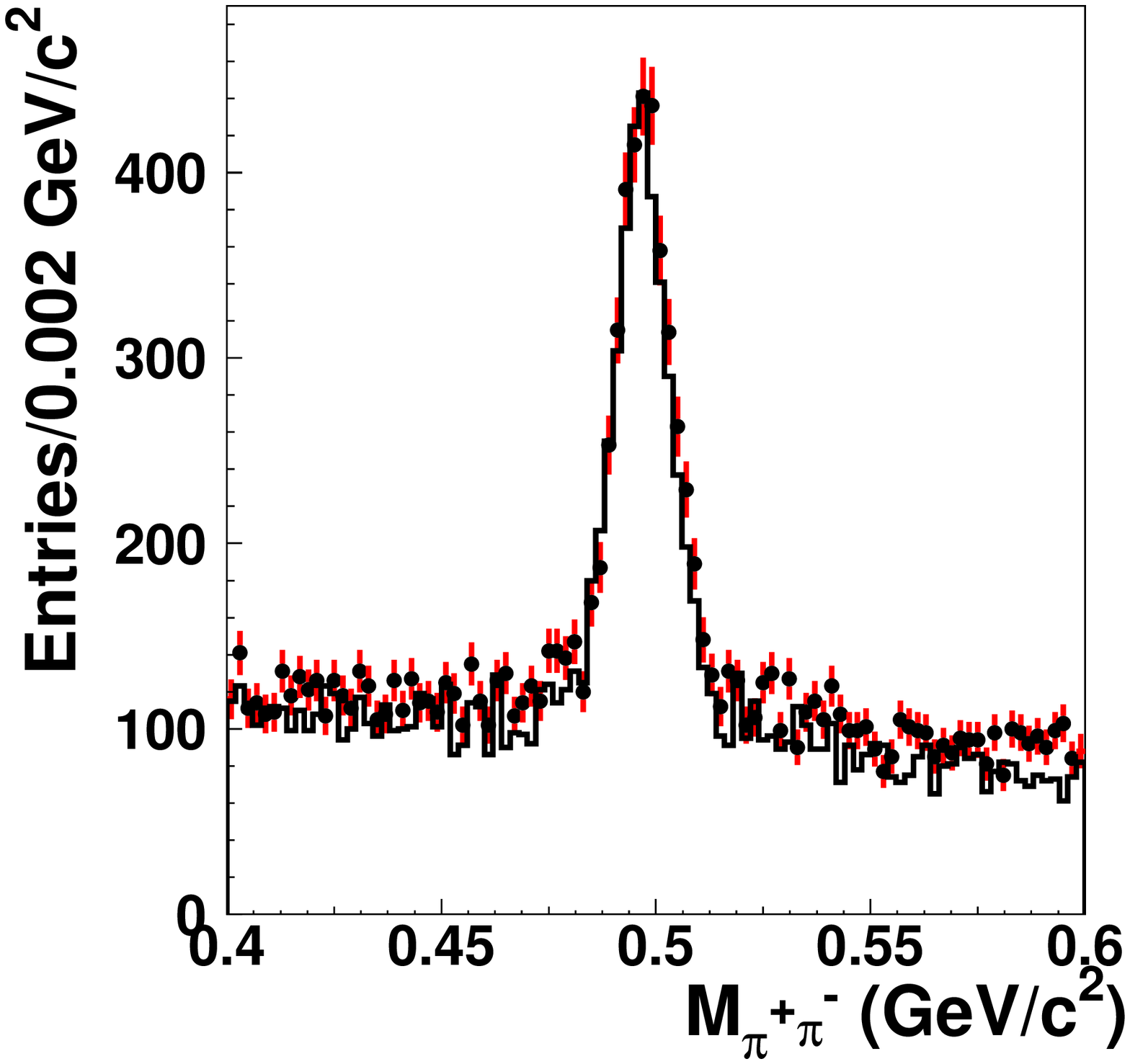}
\includegraphics[width=3.85cm,height=3.78cm]{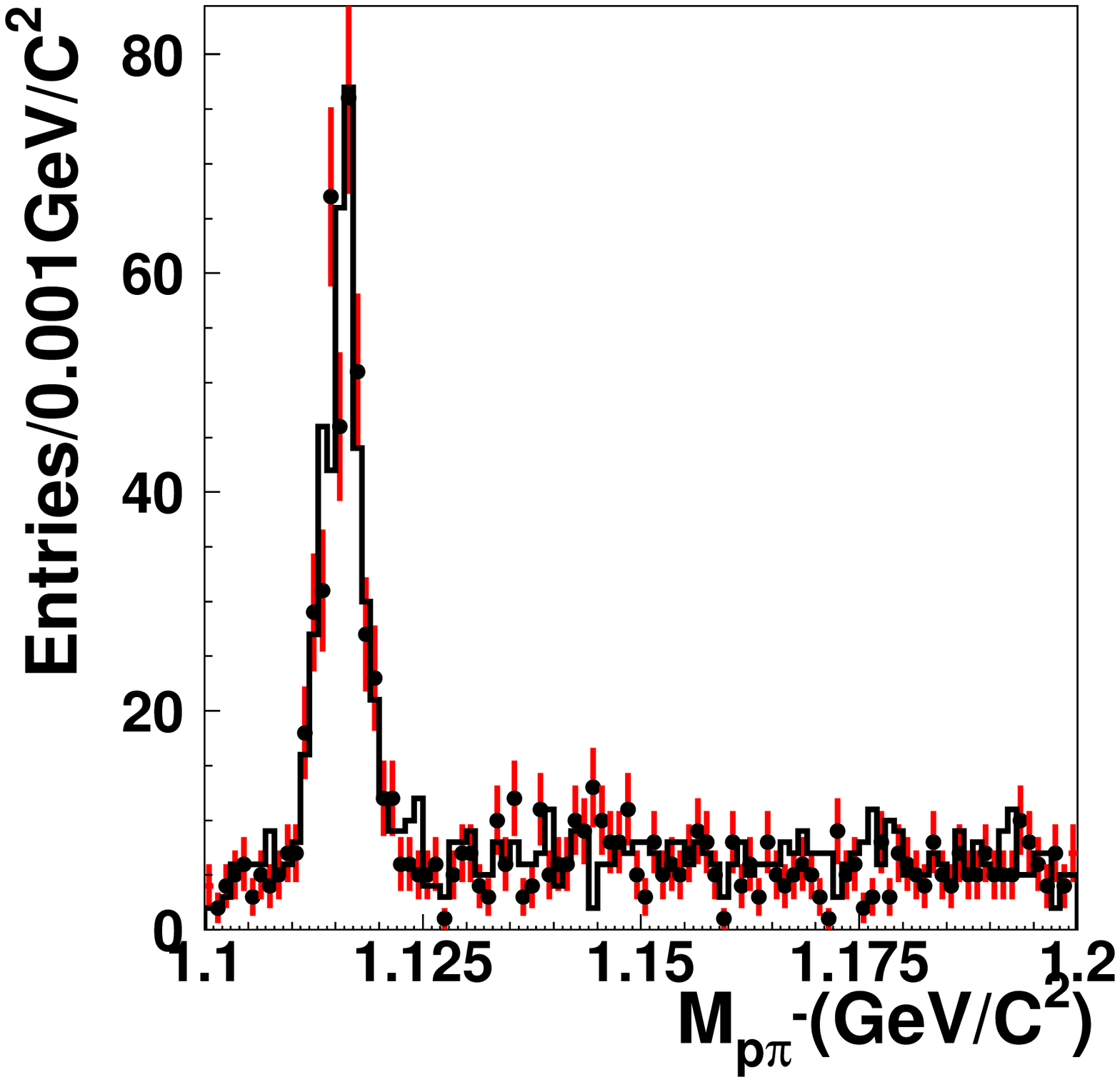}
\includegraphics[width=3.85cm,height=3.75cm]{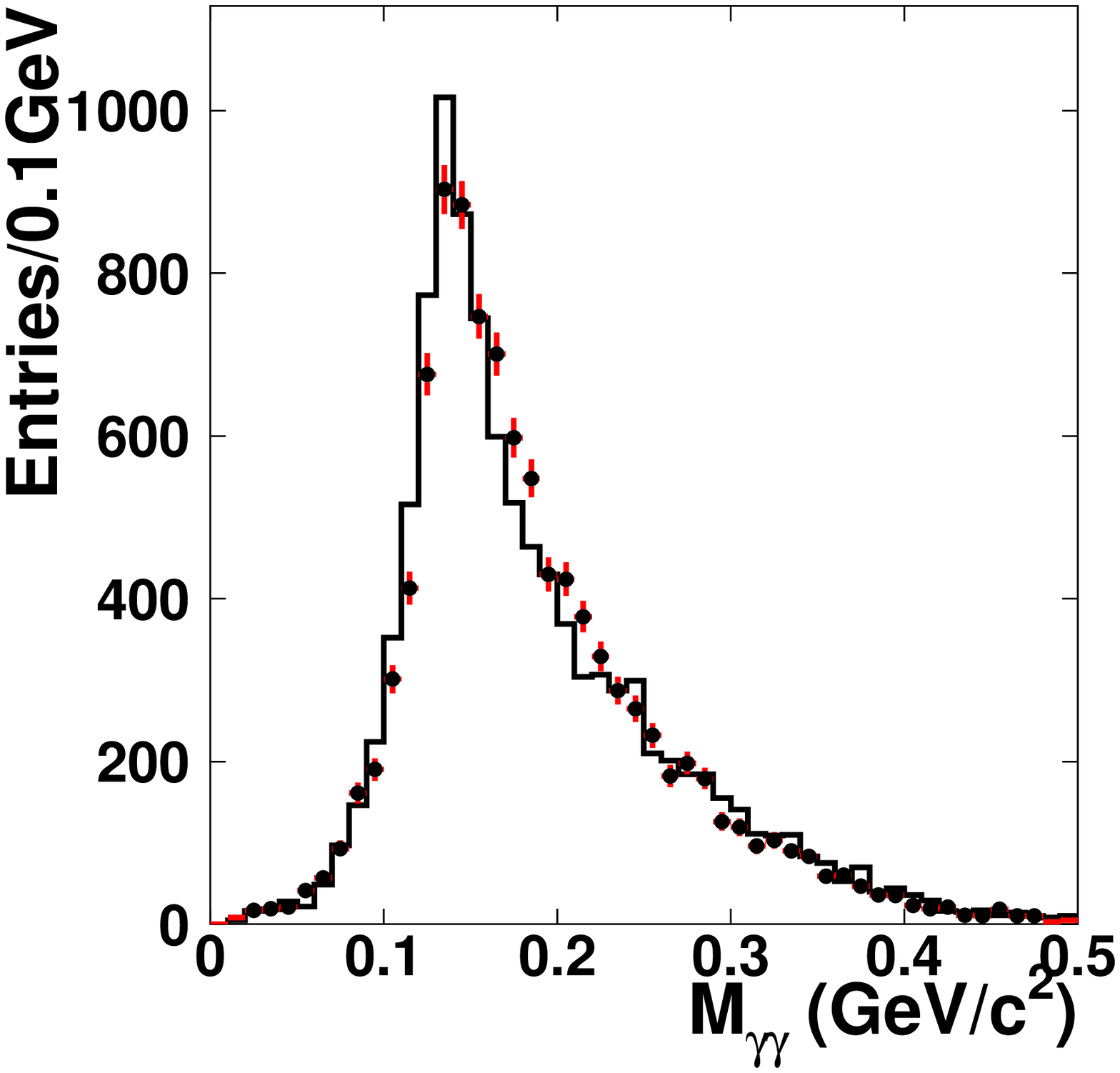}
\put(-360,95){(e)}\put(-250,95){(f)}\put(-140,95){(g)}\put(-30,95){(h)}
\caption{The normalized distributions for data (dots with error bars
) and the LUARLW MC (histograms) at 3.65 GeV with full detector
simulation: (a) multiplicity of charged tracks; (b) multiplicity of
neutral clusters; (c) momenta $p$ of charged tracks; (d)
polar-angles between charged tracks and beam direction,
$\cos\theta$; (e) deposited energies in the BSC; invariant masses of
(f) $K_S\to\pi^+\pi^-$; (g) $\Lambda\to p\pi^-$ and (h)
$\pi^0\to\gamma\gamma$ decays respectively.}\label{datalundarea}
\end{figure}

\begin{figure}[htbp]
\begin{center}
\includegraphics[width=6.5cm,height=4.8cm]{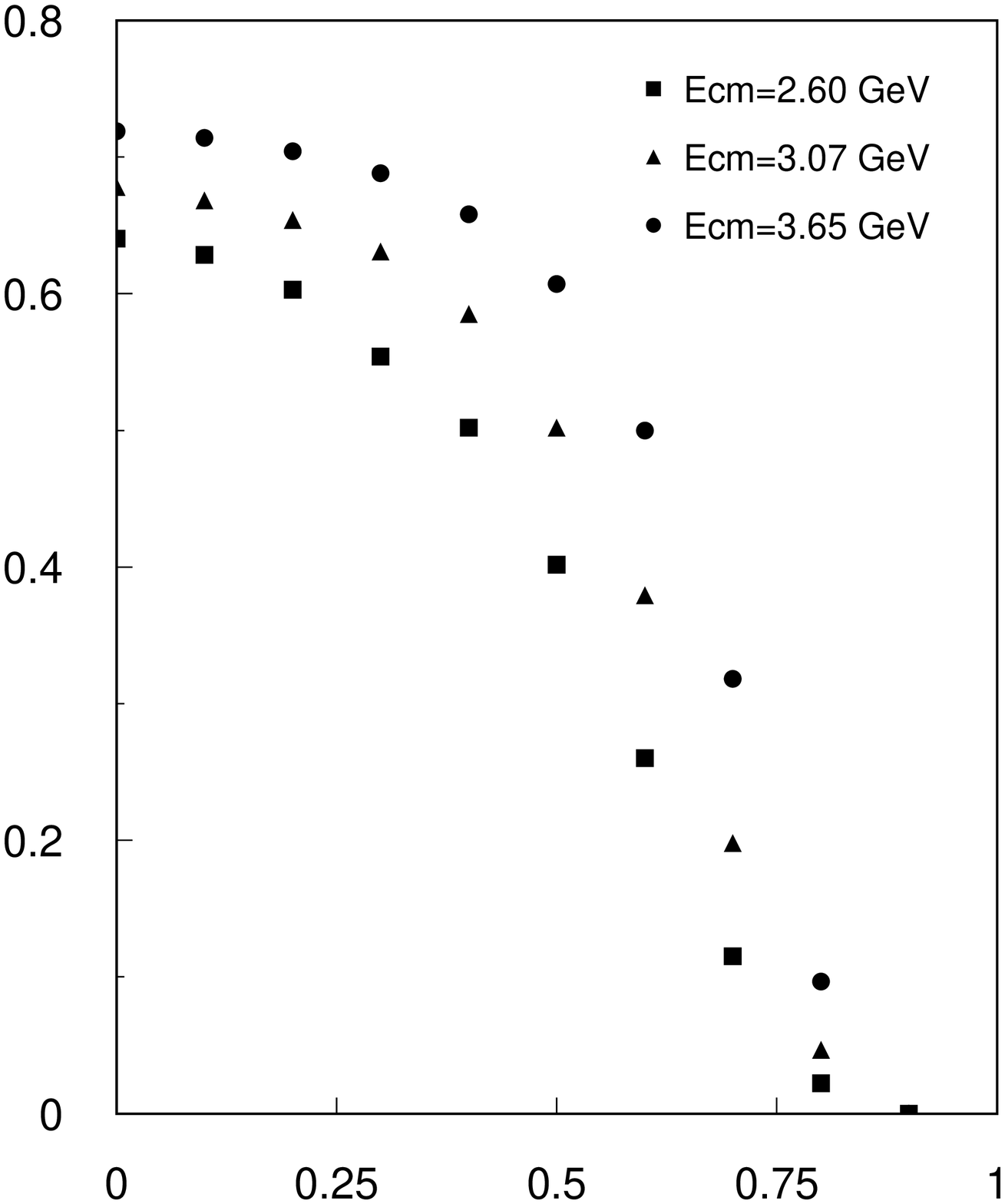}
\put(-195,115){\large $\epsilon(k)$} \put(-20,-9){\large $k$}
\vspace{-0.5cm}\caption{Radiative efficiencies $\epsilon (k)$ for
selected hadronic events with $n_{ch}\geq1$ at some values of
$k$.}\label{radeff}
\end{center}
\label{r04}
\end{figure}

\begin{figure}[htbp]
\begin{center}
\includegraphics[width=10cm,height=15cm,angle=-90]{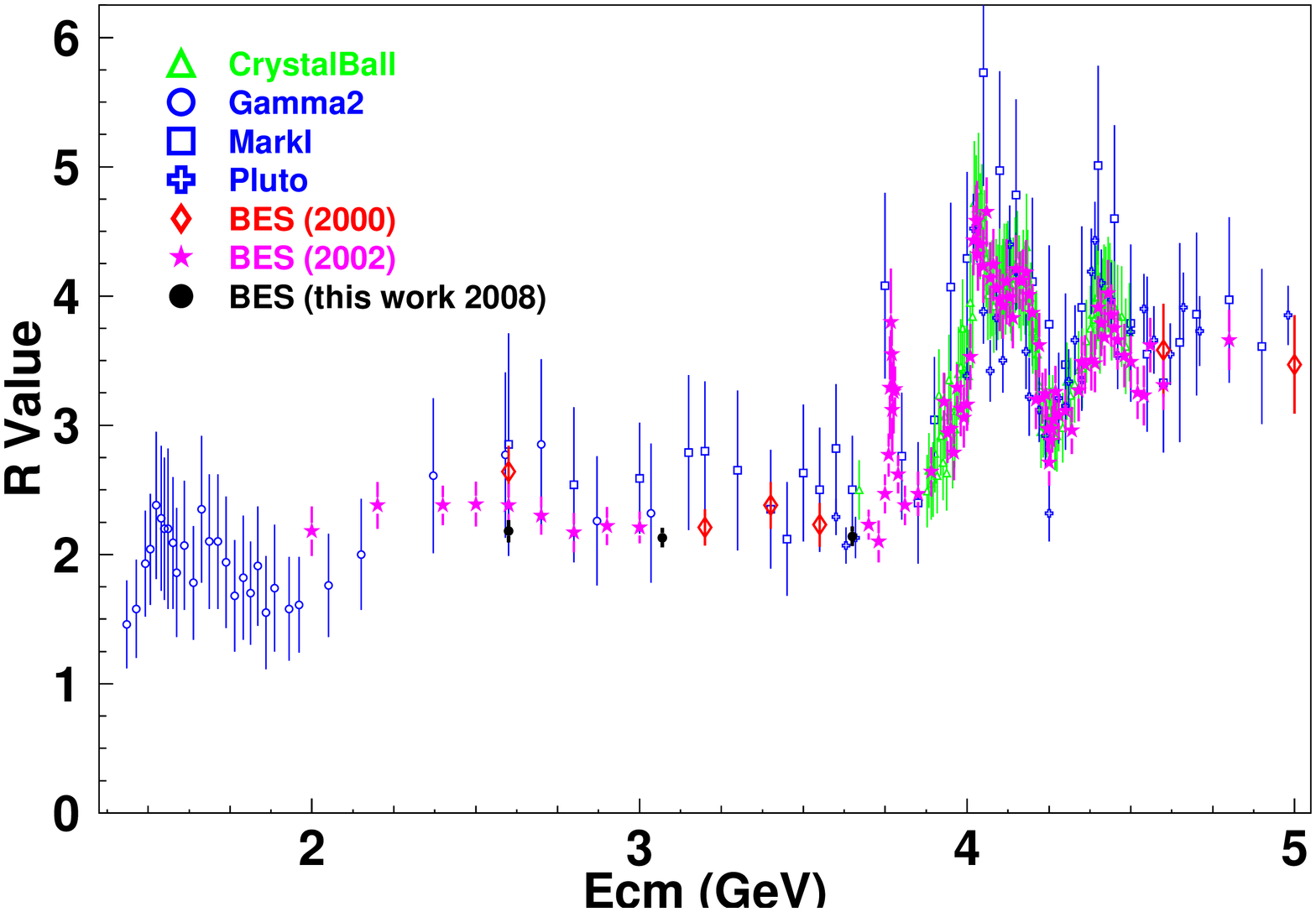}
\vspace{-1mm}\caption{$R$ values reported here together with prevous
measurements below 5 GeV.}\label{besr04}
\end{center}
\end{figure}

\end{document}